# The Visual Decoding of the "Wheel of Duality" in Consumer Theory in Modern Microeconomics: An Instructional Tool Usable in Advanced Microeconomics to Turn "Pain" into "Joy"

Seyyed Ali Zeytoon Nejad Moosavian[1]

[1] Department of Economics, North Carolina State University, Raleigh, NC, USA.

Correspondence: Seyyed Ali Zeytoon Nejad Moosavian, Department of Economics, North Carolina State University, Raleigh, NC, USA.

**Abstract**

Duality is the heart of advanced microeconomics. It exists everywhere throughout advanced microeconomics, from the beginning of consumer theory to the end of production theory. The complex, circular relationships among various theoretical microeconomic concepts involved in the setting of duality theory have led it to be called the "wheel of pain" by many graduate economics students. Put simply, the main aim of this paper is to turn this "wheel of pain" into a "wheel of joy". To be more specific, the primary purpose of this paper is to graphically decode the logical, complex relationships among a quartet of dual functions which present preferences as well as a quartet of demand-related functions in a visual manner.

## 1. Introduction

According to Cornes (2008), "dual arguments have, in recent years, become standard tools for analysis of problems involving optimization by consumers and producers." He affirms that the dual techniques have been being widely used among economists in recent decades. In his opinion, familiarity with basic duality theory is now beginning to become a necessity for graduate economics students, whether they are interested in economic theory or in empirical applications of economic theory. Cornes (2008) also adds that "the paucity of simple introductory exposition of these techniques is both surprising and disappointing." The present paper is an attempt to fill the mentioned gap, making the learning process easier for graduate economics students so that they can deeply understand and readily remember the theoretical concepts in advanced microeconomics.

Although a few other visualizations of duality in consumer theory have already been around for a while, the one that is introduced in the present paper is the most comprehensive and innovative among others which have been put forth thus far. The visualization of "wheel of duality" (WOD) introduced in this paper contains at least five more functions, decodes at least ten more connections in a visual way, and discovers several more symmetric loops. The basic idea behind this paper lies in the fact that providing one-by-one and pairwise relationships may not present the whole picture of the WOD at once, which in turn may result in building up an incomplete vision towards the essence of the duality theory. However, providing a comprehensive, visual WOD can clear up any possible confusion in this regard.

This comprehensive, visual "wheel of duality" can have two outcomes: a theoretical outcome as well as an instructional outcome. First, it gives an intuitive understanding of dual theory, which can in turn contribute to deepening students' understanding of the duality theory in advanced microeconomics, allowing them to dig further into the theory of duality. Second, this big picture can cause the complicated, circular relationships among theoretical microeconomic concepts involved in the setting of duality theory to become much more simplified and a lot more easy-to-digest and -remember for graduate economics students than ever before. Besides this, this graphical demonstration of WOD would be a practical example of utilizing visualization to improve practices in the teaching of economics. It will provide an example supporting the idea that teaching and learning economics does not have to be difficult even at a graduate level.

According to Zeytoon Nejad (2016a), "the plurality and variety of concepts, variables, diagrams, and models involved in economics can be a source of confusion for many economics students. However, reviewing the existing literature on the importance of providing visual "big pictures" in the process of learning at a college level suggests that furnishing students with a visual "big picture" that illustrates the ways through which those numerous, diverse concepts are connected to each other could be an effective solution to clear up the mentioned mental chaos." As a practical example, the present paper introduces a visual "big picture" that can be used as a valuable resource in advanced microeconomics courses. This figure mostly focuses on the ways through which the main elements of duality theory in advanced microeconomics are connected to each other, and finally introduces a holistic visual WOD that graphically demonstrates these connections. It also shows how to make transitions among these connections. In sum, this paper attempts to illustrate how one can turn a "wheel of pain" into a "wheel of joy" through a thoughtfully designed visual aid.

## 2. Literature Review

### *2.1 Usefulness of Visualization in Teaching and Learning*

Gilbert (2010) emphasizes the vital importance of representations in the process of learning by saying that "representations are the entities with which all thinking is considered to take place. Hence, they are central to the process of learning and consequently to that of teaching." Arcavi (2003) studies the role of visual representations in the learning of mathematics. Arcavi (2003) points out that "vision is central to our biological and socio-cultural being" (Arcavi, 2003, p.213). Therefore, as biological and as socio-cultural beings, we are encouraged and aspire to 'see' not only what comes 'within sight', but also what we are unable to see." He then refers to a quote from McCormick et al. (1987) stating that "visualization offers a method of seeing the unseen" (Arcavi, 2003, p.216).

In her book called "Teaching at its Best", Nilson (2010) states that structure is so key to how people learn. It has such far-reaching implications for teaching. She believes without structure there is no knowledge. She says "information" is nowadays available everywhere. However, what it is not so available everywhere is organized bodies of "knowledge". She defines knowledge as a structured set of patterns that we have identified through observation. She argues that students are not stupid; they are simply novices in the discipline, who do not see the big picture of the patterns, generalizations, and abstractions that experts recognize so clearly (Arocha & Patel, 1995; DeJoneg & Ferguson-Hessler, 1996). She warns instructors that without such a big picture, students face another learning hurdle in addition to other hurdles they may already have.

It has been known that the human mind processes, stores, and retrieves knowledge not as a collection of facts, but as a logically organized whole, a coherent conceptual framework, with interconnected parts. Without having a structure of the material in their heads, students fail to comprehend and retain new material (Anderson, 1984; Brandsfor et al., 1999; Svinicki, 2004). The kind of deep, meaningful learning that moves a student from novice toward expert is all about acquiring the discipline's hierarchical organization of patterns, its mental structure of knowledge (Anderson, 1993; Royer, Cisero & Carlo, 1993). "Only then will the student have the structure needed to accumulate additional knowledge" (Nilson, 2010, p.6). As Nilson (2010) reports, according to Kozma et al. (1996), since the chances are very slim that students will independently build such cognitive schemata in a semester or two of casual study, it is wise instructors' task to furnish their students with relevant structure of the associated discipline with valid, ready-made frameworks for fitting the content.

To conclude the literature reviewed here, it should be noted that there is a huge potential with providing visual "big pictures" of complex theoretical economics subjects that economics instructors can take advantage of so as to improve the quality of teaching and learning of economics. It seems that this potential capacity has not yet been fully employed to solve some of the issues with the teaching of economics. The present paper is an effort to fill this gap in consumer theory by visualizing the subtleties and complexities of the employment of the theory of duality in the setting of consumer theory.

### *2.2 Multiple Instances of Providing Visual "Big Pictures" in Economics and Its Related Sciences*

Before going any further, I find it more helpful to first cite multiple well-known, widely-used, and innovative visual "big pictures" in economics and its related disciplines such as statistics.

Speaking of a best practice of a visual "big picture in the area of statistics, I would like to cite Leemis and McQueston (2008) that provide an excellent example of a "big picture" for probability distribution families and their relationships. This "big picture" of distributions not only illustrates the ways through which the distributions are connected, but it also gives some details in a notational form to make those relationships clear to audiences. Speaking of an instance of a visual "big picture" in the area of microeconomics, Zeytoon Nejad (2016a) provides a fine example of a visual "big picture" of how general equilibrium of macroeconomics is formed in the IS/LM/AS/AD framework. This visual "big picture" can be employed as a good resource in intermediate macroeconomics classes. This figure presents twenty-seven commonly-discussed macroeconomic diagrams in the intermediate macroeconomics course, and gives little detail on some of the macroeconomics diagrams, aiming at helping students to get the whole picture at once on a single piece of paper. Speaking of microeconomics, Snyder and Nicholson (2012) give a notable example of a visual "big picture" linking several demand-related concepts in a single picture (Figure 1). When these concepts are introduced one by one, separately, and without using any visualization, they seem to be complicated at first glance to most of the students who have newly started dealing with these ideas. Nevertheless, providing the students with such an explanatory visual "big picture" helps them readily figure out how those ideas and concepts are related and linked to one another.

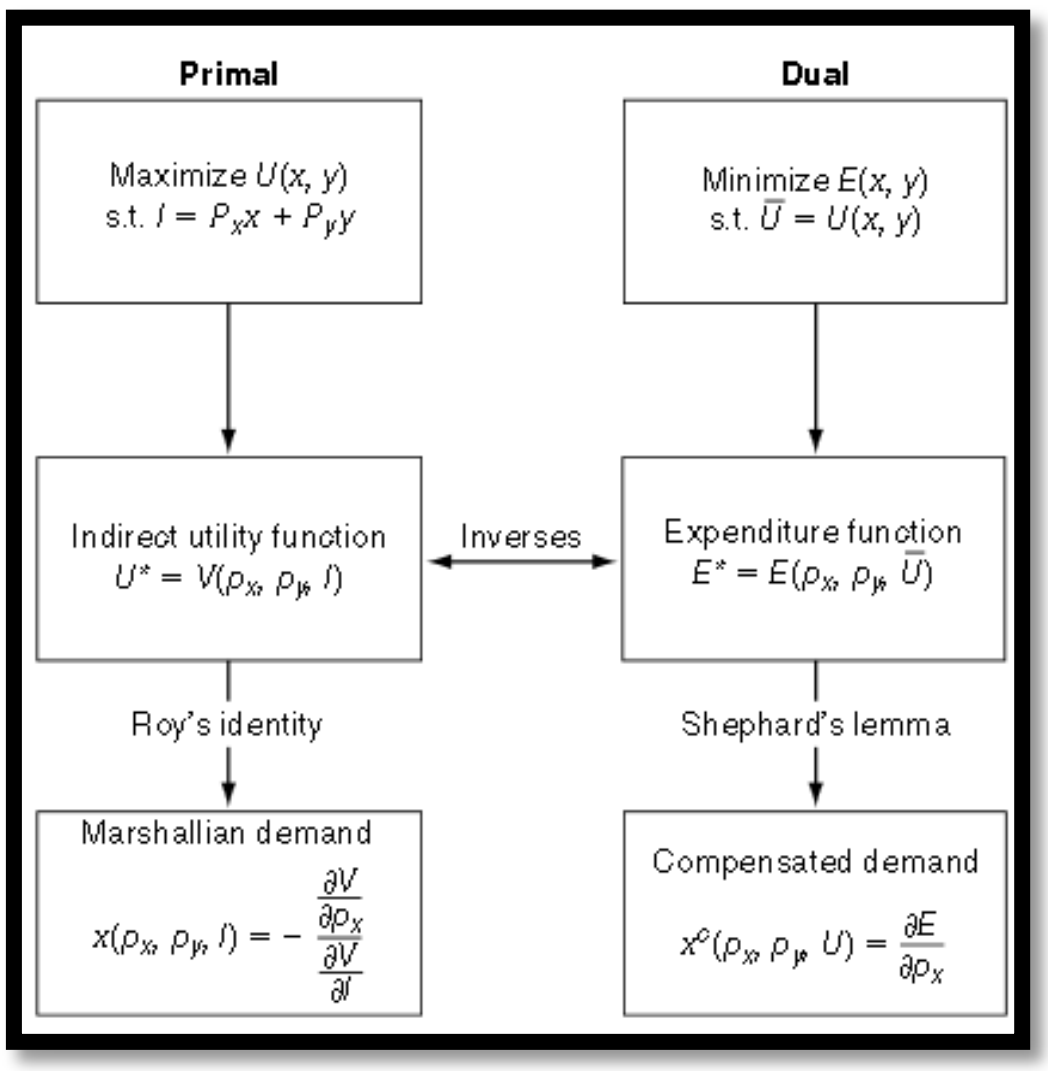

Figure 1. A Big Picture Linking the Relationships among Demand-related Concepts

By: Snyder and Nicholson (2012)

Nonetheless, the present paper claims that there is a lot more going on in the context of the theory of duality in consumer theory. That is, in the remaining sections, this paper will try to make the point that the above-presented visual "big picture" is a good, but not sufficient, visual aid to describe all the complex, circular linkages among the theoretical microeconomic concepts involved in the setting of the duality theory. In the next section, the main discussion of the paper will be offered, during which so many of other mathematical aspects involved in the duality theory application in consumer theory will be revealed. The structure of the next section is such that firstly a brief discussion is made about the duality theory from a mathematical point of view. Then, the applications of the theory of duality in the context of consumer theory are discussed in some details. Afterwards, the comprehensive visual WOD in consumer theory is built up step by step, and then its specifications, subtleties, and features are explained in brief. Finally, some considerations on how to use it when teaching a course will be presented.

## 3. Main Discussion

### *3.1 Duality Theory in Mathematics and Its Applications in Consumer Theory*

Duality is an extensive mathematical topic. Hence, introducing all of its technical aspects in details is beyond the scope of this paper. This section is, instead, to discuss the mathematical dimensions of the duality theory in brief first. Immediately after, I will approach this theory from a microeconomic angle, and will focus more on the application of the theory of duality in consumer theory. Thus, I will suffice to give a short mathematical explanation of the duality theory.

According to the duality principle, optimization problem may be viewed from either of two perspectives, from the *primal*-problem viewpoint or from the *dual*-problem viewpoint. The solution to the dual problem provides a lower bound to the solution of the primal problem. In general, however, the optimal values of the primal and dual problems need not be equal. The difference between the optimal value of the primal (p*) and the optimal value of the dual problems (d*) is called the *duality gap* (p* - d*). For convex optimization problems and when strong duality conditions hold, the duality gap is zero under a constraint qualification condition (Boyd and Vandenberghe, 2004).

Mathematically speaking, the following problems (figure 2) are the generic forms of the primal and dual problems for a normal linear programing problem. An excellent example of the application of the mathematical theory of duality in economics is the use of duality theory in the context of consumer theory, in which for any utility maximization problem, there exist a corresponding expenditure minimization problem. Figure 3 provides an intuitive illustration of the duality theory in modern consumer theory, in which utility maximization problem has been located on the left-hand side diagram and expenditure minimization problem has been located on the right-hand side diagram.

**Primal Problem**

$$\begin{aligned} \max\ z &= c_1x_1 + c_2x_2 + \dots + c_nx_n \\ \text{s.t.}\quad & a_{11}x_1 + a_{12}x_2 + \dots + a_{1n}x_n \le b_1 \\ & a_{21}x_1 + a_{22}x_2 + \dots + a_{2n}x_n \le b_2 \\ & \vdots \\ & a_{m1}x_1 + a_{m2}x_2 + \dots + a_{mn}x_n \le b_m \\ & x_j \ge 0 \quad (j = 1, 2, \dots, n) \end{aligned}$$

**Dual Problem**

$$\begin{aligned} \min\ w &= b_1y_1 + b_2y_2 + \dots + b_my_m \\ \text{s.t.}\quad & a_{11}y_1 + a_{21}y_2 + \dots + a_{m1}y_m \ge c_1 \\ & a_{12}y_1 + a_{22}y_2 + \dots + a_{m2}y_m \ge c_2 \\ & \vdots \\ & a_{1n}y_1 + a_{2n}y_2 + \dots + a_{mn}y_m \ge c_n \\ & y_i \ge 0 \quad (i = 1, 2, \dots, m) \end{aligned}$$

Figure 2. The Generic Forms of the Primal and Dual Problems for a Normal Linear Programing Problem

Adopted from Winston (1994), Operations Research: Applications and algorithms

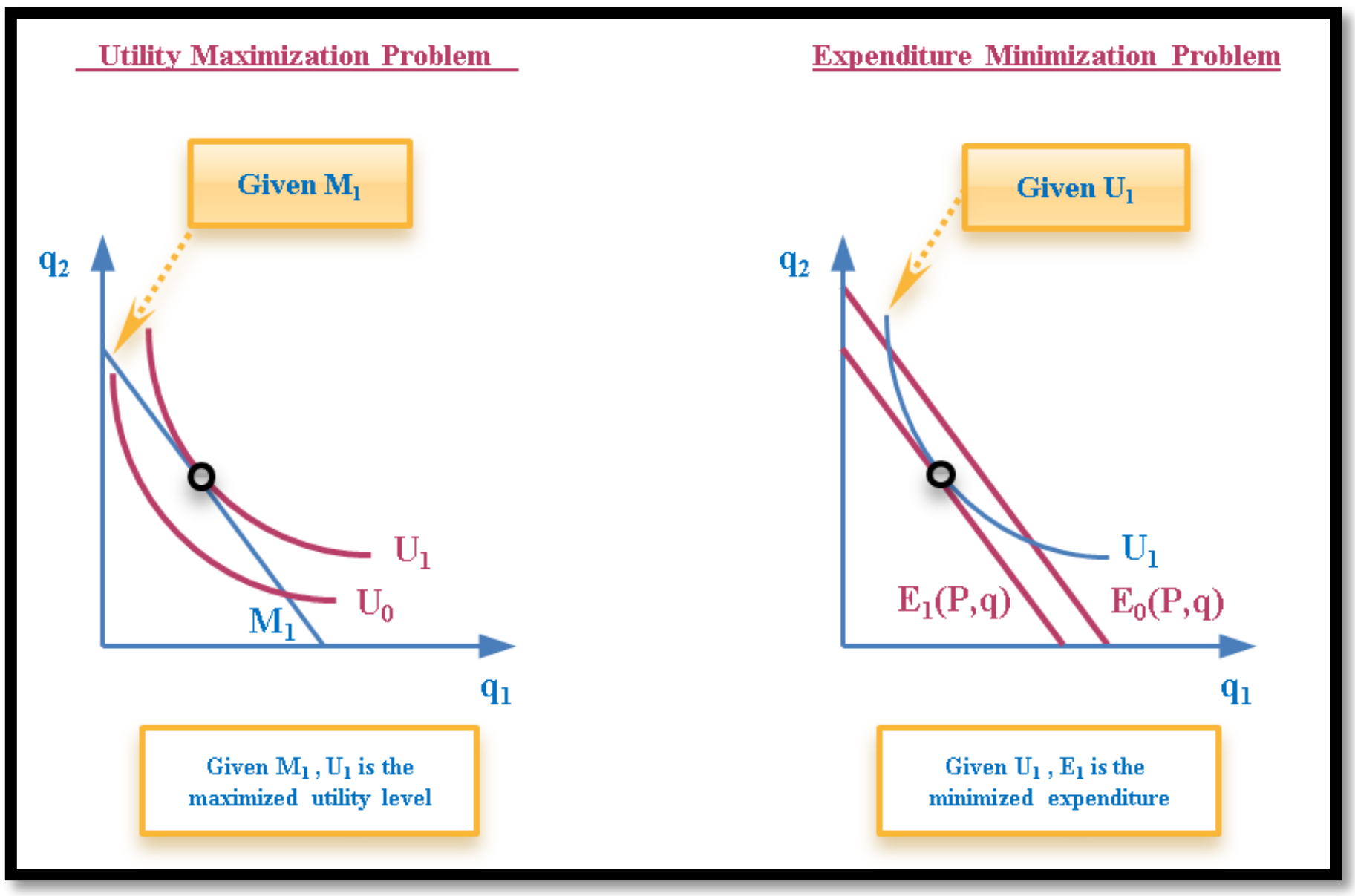

Figure 3. An Intuitive Illustration of the Duality Theory in Modern Consumer Theory

Theoretically, there are eight axioms that are usually assumed to be satisfied by preferences when considering consumer choice in consumer theory. This way, we can order preferences in a consistent way, define utility functions, and make the duality theory work well in the context of consumer theory. Based on these eight axioms, individuals' preferences must be reflexive, complete, transitive, continuous, non-satiable, convex, (preferably) strictly convex, and differentiable. Among the aforementioned characteristics, the first three comprise the essence of rationality in an individual consumer behavior. The first four are sufficient to define a working utility function. In order for the theory of duality to work meaningfully in the context of consumer theory, preferences must satisfy the first six axioms. The last two greatly simplify the exposition of the duality theory (Cornes, 2008). The following figure summarizes the above-mentioned discussion.

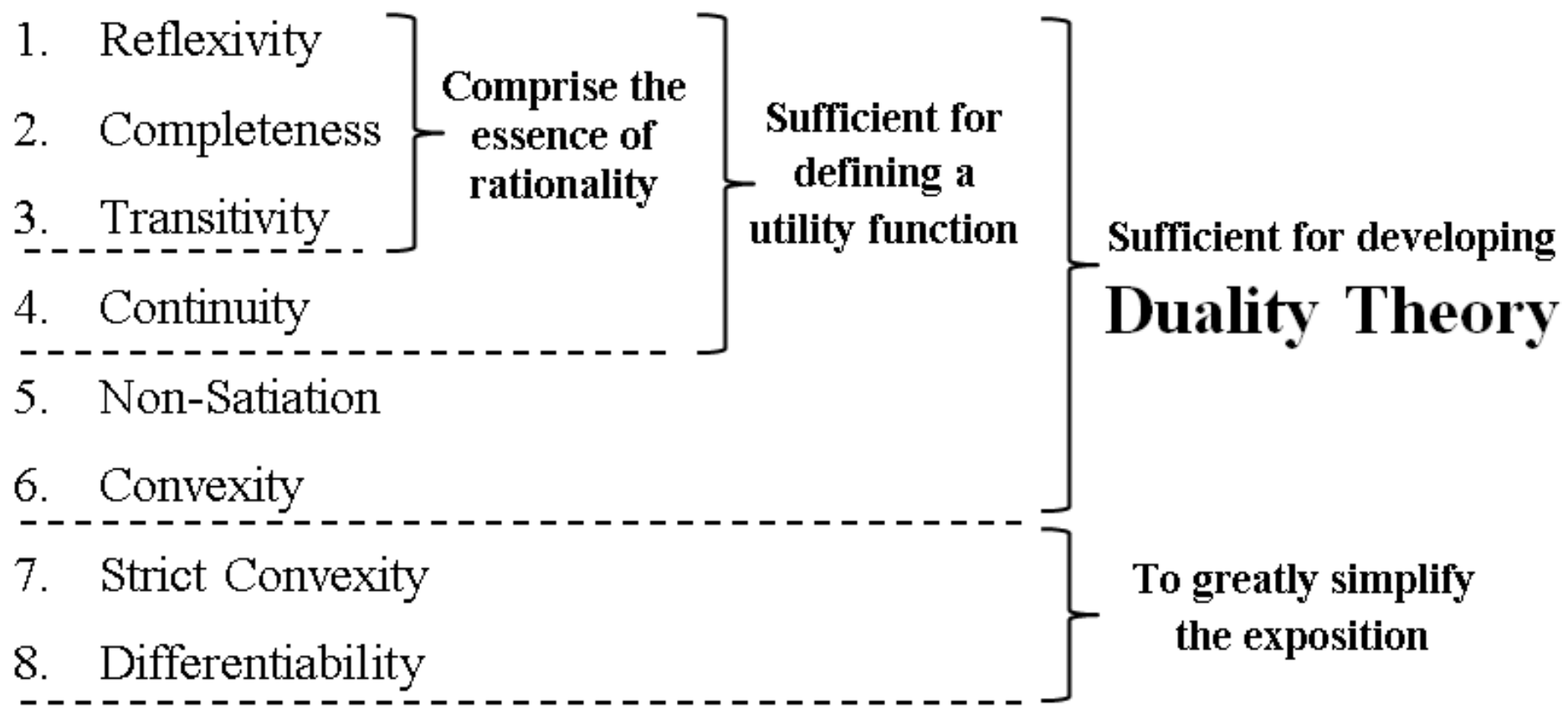

Figure 4. Axioms of Preferences

In principle, there are 4 alternatives possibilities for describing the preference ordering of a consumer by a mathematical function: Direct Utility Function (DUF), Indirect Utility Function (IUF), Expenditure Function (EF), and Distance Function (DF).[1] Each of these functions gives rise to one type of demand function, namely Hotelling-style Inverse Demand Function (HIDF)[2], Marshallian Demand Function (MDF), Hicksian Demand Function (HDF), and Antonelli-style Inverse Demand Function (AIDF)[3], respectively. The following figure summarizes all the connections outlined above.

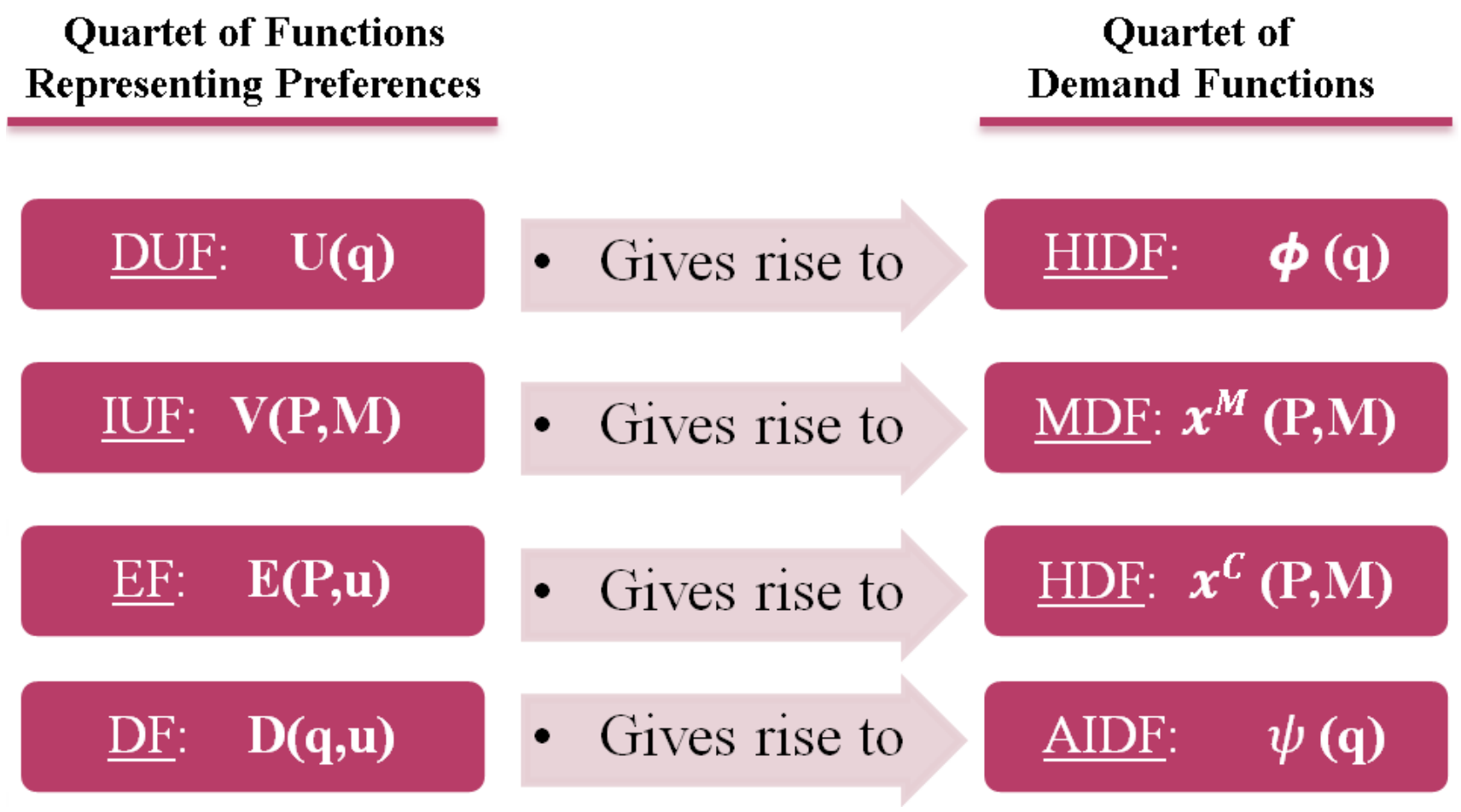

Figure 5. The Quartet of Functions Reflecting Preferences vs. the Quartet of Demand Functions

*3.2 Step-by-Step Design of the Visual Wheel of Duality in Consumer Theory*

Now, the goal of the rest of this section is to show how these eight functions are linked to each other in a visual manner. In so doing, it will be helpful to first clearly classify various types of relationships among the functions existing in the wheel of duality (WOD). In total, there are four types of relationships in the visual WOD which is to be introduced in this paper. The first type of relationship among these functions is the case in which two functions are dual of each other. The relationship between DUF and IUF as well as the relationship between DF and EF are of this type. Mathematically speaking and put simply, when one function is the dual function of another, it practically means that one function can be derived from the other.[4] For instance, the following problems show the reason why DUF and IUF are called dual of each other.[5]

[1] It is important to note that it is not the aim of the present paper to discuss all the technical and mathematical details or applications of the functions being examined. Instead, the present paper assumes that readers already have the needed background knowledge about the subject matter to some extent. Thus, the paper primarily focuses on the types of relationships among those functions.

[2] The relationship between the market price and the partial derivatives of DUF is often referred to as the Hotelling-Wold identity, in acknowledgment of two economists who were among the first to drive it (Cornes, 2008). Hence, normalized, inverse MDF is here referred to as the Hotelling-style Inverse Demand Function (HIDF in short).

[3] According to Cornes (2008), although normalized, inverse HDFs are not as well known as HDFs themselves, they have a respectable history. The Italian economist Antonelli was the first one to discuss normalized, inverse HDFs as long ago as 1886. Hence, normalized, inverse HDF is here referred to as the Antonelli-style Inverse Demand Function (AIDF in short).

$$\boldsymbol{U}(\boldsymbol{q}) \equiv \min_{P}\{\boldsymbol{V}(\boldsymbol{P},\boldsymbol{M})|P.q \geq M\}$$

$$\boldsymbol{V}(\boldsymbol{P},\boldsymbol{M}) \equiv \max_{q}\{\boldsymbol{U}(\boldsymbol{q})|P.q \leq M\}$$

The same sort of relationship applies to the relationship between DF and EF.

$$\boldsymbol{D}(\boldsymbol{q},\boldsymbol{u}) \equiv \max_{P}\{P.q|\boldsymbol{E}(\boldsymbol{p},\boldsymbol{u}) = 1\}$$

$$\boldsymbol{E}(\boldsymbol{p},\boldsymbol{u}) \equiv \min_{q}\{P.q|\boldsymbol{D}(\boldsymbol{q},\boldsymbol{u}) = 1\}$$

Figure 6 depicts the first step in building up the comprehensive visual WOD that the present paper is after. As shown below, the utility maximization problem is defined as the primal problem and the expenditure minimization problem as the dual problem.

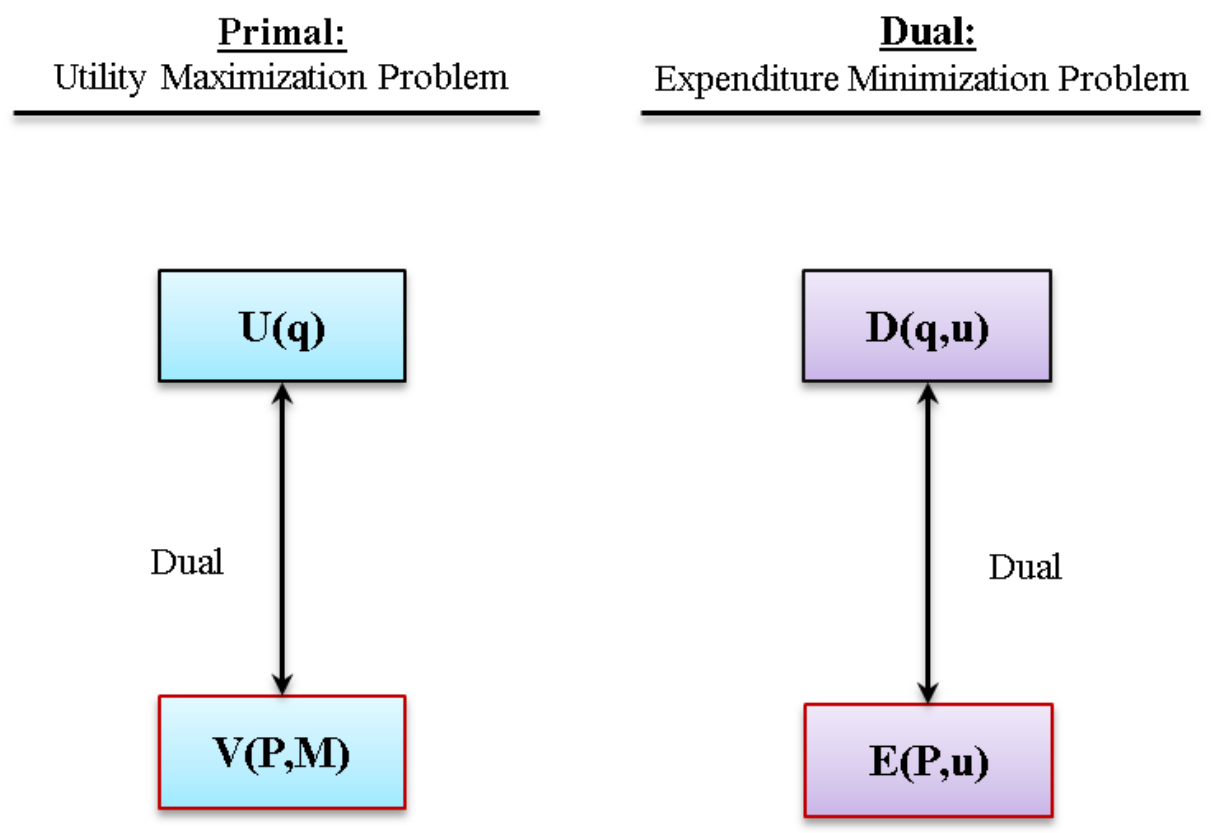

Figure 6. Dual Functions under the Primal and Dual Problems

A second type of relationship between the functions existing in the WOD is being mathematical inverse functions of each other. That is, in this case, two functions are mathematical inverse of one another. The relationships between the following pairs of functions are of this kind: DUF and DF, HIDF and MDF, AIDF and HDF, and finally IUF and EF. The following figure demonstrates how the said functions are fit into the comprehensive, visual WOD that is to be introduced in the present paper.

[4] This is just a working definition of a dual function. There are more complex, technical aspects to the definition of a dual function from a mathematical point of view. To obtain further information on the mathematical definition of a dual function, you can see mathematical textbooks on the theory of duality and convex optimization such as Boyd and Vandenberghe (2004).

[5] It is also important to clearly differentiate between the two expressions “the dual problem”, which is essentially an alternative setup for the primal problem, and “dual functions”, which are the functions that can be derived from each other, primarily through optimization. Obviously, dual functions can appear under either of the primal or dual problem.

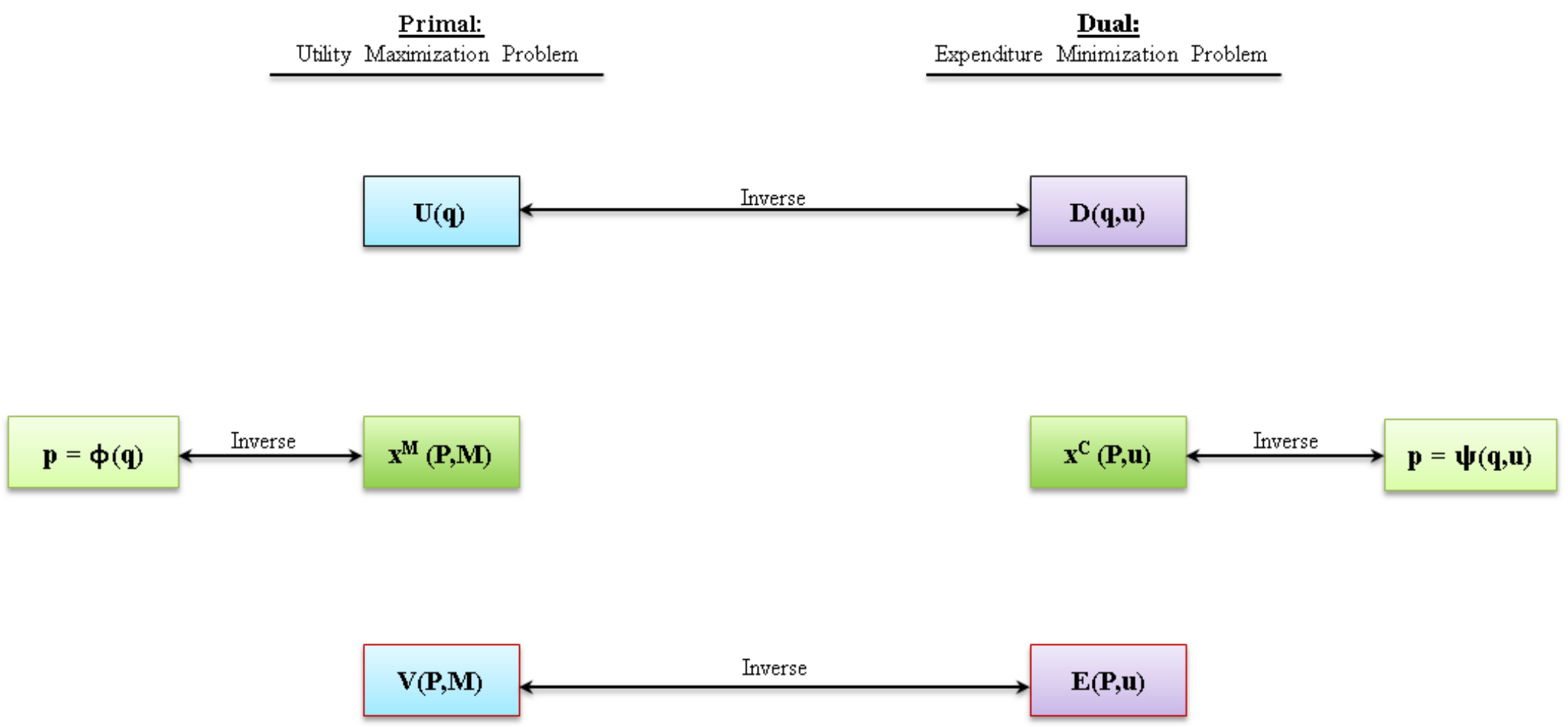

Figure 7. Inverse Functions under the Primal and Dual Problem in the WOD

A third type of relationship that the components of the WOD presented here may have to one another is the case in which two functions are counterparts of each other under primal and dual problems. For instance, MDF, which is obtained by solving the primal problem, is the counterpart of HDF, which is obtained by solving the dual problem, and vice versa. The reason why they are called counterparts is because either of these functions represents the optimal quantities under its corresponding problem. With the same reasoning, HIDF and AIDF are counterparts of each other, since they are optimal prices under the primal and dual problem, respectively. Figure 8 depicts how these counterparts fit into the WOD.[6]

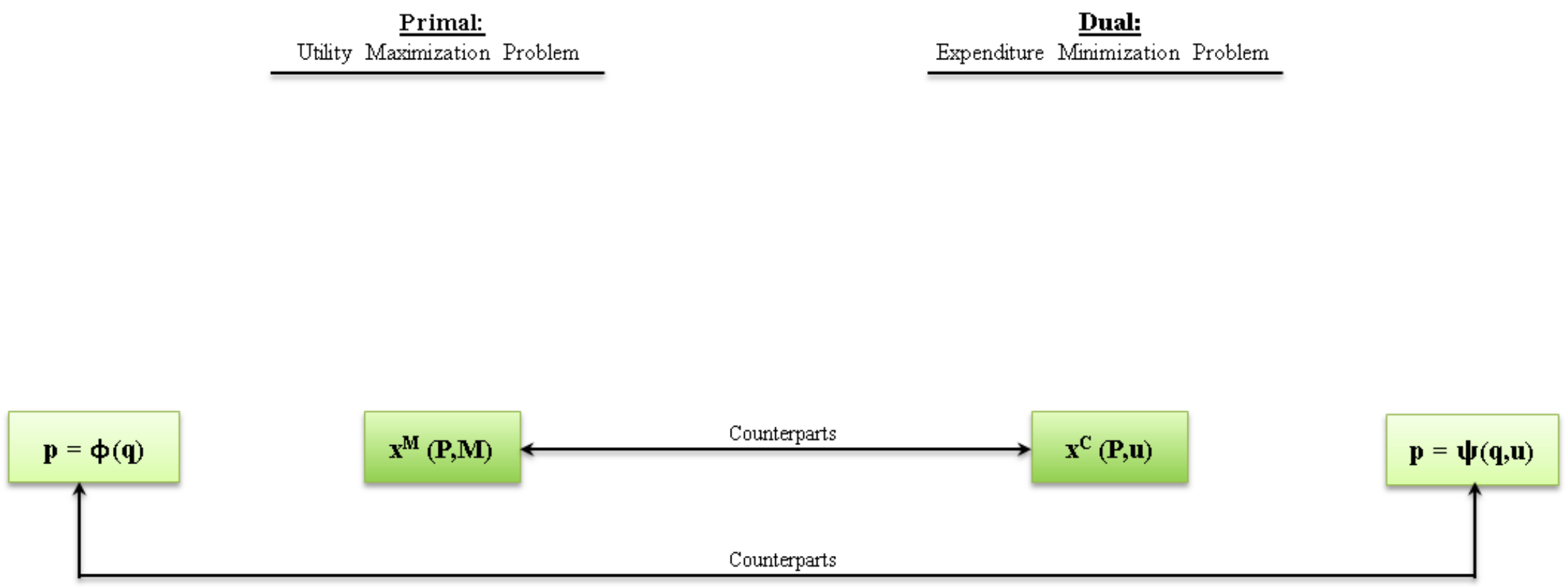

Figure 8. Counterpart Functions under the Primal and Dual Problems

A fourth type of relationships in the WOD is a "derivative relationship" through which one function is derived from another through a mathematical operation, equation, identity, or lemma. For instance, the pairs of DUF and HIDF, or IUF and MDF, or EF and HDF, or DF and AIDF have this kind of relationship to each other, in the sense that the former functions give rise to the latter ones as outlined already in figure 5. Besides this, there are many other derivative relationships in the WOD that can readily be identified by looking at the WOD, such as the relationship between DUF and MDF in which case MDF is derived from maximizing DUF subject to the BC.

Figure 9 exhibits the comprehensive version of the WOD which includes all the mathematical formulations and operations needed to make theoretically meaningful transitions among different functions existing in the WOD. As shown in figure 9, the first cell on the upper left side shows the setup of the primal problem in which DUF is the objective function to be maximized and the BC is the constraint of the problem. Then, the problem is to maximize DUF subject to the BC. On the other side of the WOD and under the dual problem, the first cell on the upper right hand side shows the setup of the dual problem in which the “expenditure amount function” (EAF)[7] is the objective function and DF when is equal to 1 (D(q,u)=1) is the constraint of the minimization problem. Then, the problem is to minimize EAF (i.e. E(P,q)=P.q) subject to D(q,u)=1. It is important to note and remember that the variables P, p, q, $x^M$, and $x^C$ (which represent vectors of prices, “normalized” prices, quantities consumed, Marshallian demand quantity, Hicksian demand quantity, respectively) are all “vectors” throughout the WOD, not just scalars. Therefore, it makes more sense to think of them as $P_i$, $p_i$, $q_i$, $x^M_i$, $x^C_i$ where i = 1 ,…, n and n is the number of commodities under study.

[6] In some sense, it could be said that the budget constraint (BC) in the primal problem and the expenditure amount function (EAF) in the dual problem are counterparts of each other.

In the primal problem, preferences (direct utility function) are located in the objective function, while in the dual problem preferences (distance function) are placed in the constraint. It is also important to notice that preferences are situated in DUF, IUF, DF, and EF. Each of these functions is essentially a single function containing all preferences over the commodities under study. They are in fact an abstract form of preferences. On the contrary, each of demand and inverse demand functions is indeed an extensive form of preferences providing possibly a system of equations (i.e. a system of demand functions), each of which represents a demand function for one of the commodities existing in the related preference function. In sum, each of DUF, IUF, DF, and EF is a single function representing preferences over all the commodities under study, while each of HIDF, MDF, AIDF, and HDF could be a system of equations. The number of equations in each of these systems is equal to the number of FOCs of the related optimization problem. Hence, it makes more sense to put a plural “s” at the end of their acronyms as HIDFs, MDSFs, AIDFs, and HDFs implying that each of them alone is the representative of a whole system of equations.

Figure 9 summarizes all the relationship types introduced above. It also provides all the operations, equations, identities, and lemmas that help us make the aforementioned transitions. These operations, equations, identities, and lemmas are as follows: Lagrangian, mathematical substitution, mathematical inversion, price normalization (through dividing prices by income), Hotelling-Wold Identity (called H-W Id. In the visual WOD), Antonelli equations (called Antonelli), Roy’s identity (called Roy’s Id.), Shephard lemma (called Shephard), and Slutsky equation (called Slutsky). For a full list of the symbols and notations employed in the visual WOD, you can see appendix 1.

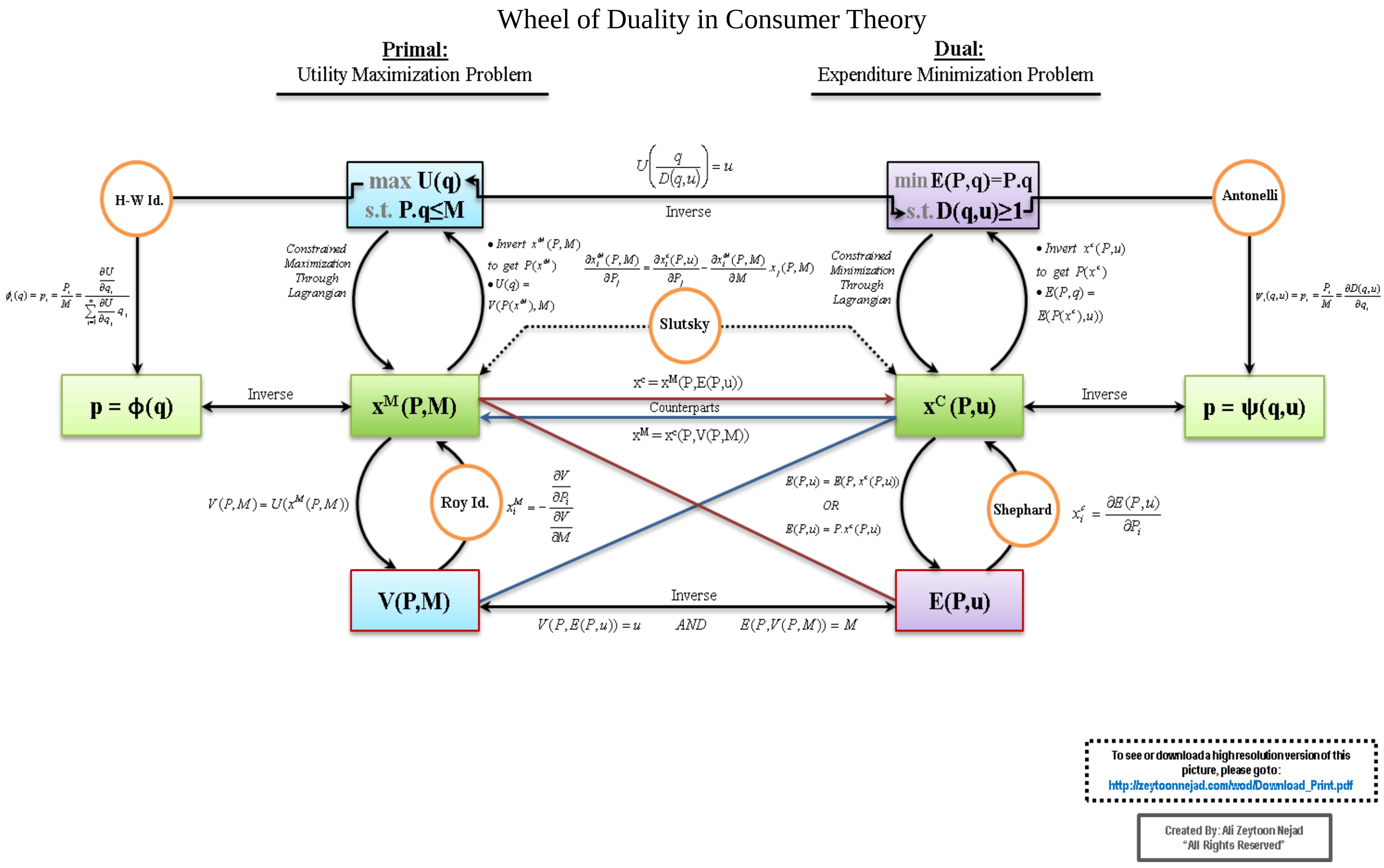

Figure 9. Wheel of Duality in Consumer Theory in Modern Microeconomics

[7] It is important to distinguish between “the” expenditure function (which called EF in short here), and expenditure amount function (which is called EAF in short here). EF is in fact *E(P,u)*, whose arguments are *P* and *u*. EAF is indeed *E(P,q)*, whose arguments are *P* and *q*.

Now that the eight main “destinations” of the WOD have been covered, the aim of this section of the paper is to explain all the “transitions” possible to make in the WOD step by step. It will be reasonable to start with the primal problem and DUF which are more well-known among economists. The most commonly-used system of demand functions among economists is the system of MDFs. Quite often, this system of demand functions is derived through mathematical maximization of DUF subject to BC either through Lagrangian or through substitution.

In order to derive MDFs, alternatively, one can first use the Hotelling-Wold identity in order to obtain a system of HIDFs from DUF as outlined in the visual WOD. This will result in a system of equations in which normalized prices are expressed as functions of quantity bundles. This transition can be employed when one is interested in expressing normalized prices as functions of quantities. Then, this system has to be inverted and its price normalization needs to be undone so that we can transition from HIDF to MDF, which gives us an indirect approach to obtaining MDFs from DUF. In order to transition back from MDFs to DUF, one needs to first list up the inverse MDFs (i.e. prices as functions of quantities) and then substitute them back into IUF to end up with DUF.

In transitioning from MDFs to IUF, one needs to simply substitute the system of MDFs into the DUF in order to get IUF. Conversely, we can take advantage of the Roy’s identity in order to transition from IUF to MDFs. EF is essentially the mathematical inverse of IUF, in which M and V are renamed as E and u, respectively. In order to make transition between these two function one can simply make use of the two equations introduced in the visual WOD under the line that connects these two to each other.

Under the dual problem, one can solve the minimization problem introduced in the WOD and thereby obtain HDFs from DF and EAF. Alternatively, this transition can be made indirectly through first obtaining AIDFs from DF by using Antonelli equation and then inverting the system of AIDFs and also undoing its price normalization so as to obtain the system of HDFs. For transitioning back from HDFs to EAF, one needs to first derive the inverse HDFs[8] (i.e. prices as functions of quantities) and then substitute them back into EF in order to end up with EAF. Additionally, for transitioning from HDFs to EF, one can readily use the routine substitution of HDFs into EAF. For the other way around, one can easily take advantage of the well-known result in microeconomics which is called Shephard’s lemma.

There are two additional possible transitions in the WOD that can be used if the needed equations are available and known. The first one is an alternative direct transition from IUF to MDFs through which IUF is substituted in its corresponding system of HDFs in place of their “u” arguments. Then, the result will be the corresponding system of MDFs. As described above and exhibited in the WOD, this transition is parallel to the Roy’s identity, and provides us with the same service, but it works quite easier if we already have HDFs as known equations. The second additional transition is an alternative direct transition from EF to HDFs through which EF is substituted into its corresponding system of MDFs in place of their M arguments. This will results in the corresponding HDFs. As outlined above and shown in the WOD, this transition is parallel to the Shephard’s lemma, and provides us with the same service as Shephard’s lemma does, but it is more convenient to use in cases where we already have MDFs as known equations.

There are two remaining points that are not explained yet: a transition and an equation. One final transition that one can make in the WOD is the transition from DUF to DF. These two functions are in fact mathematical inverses of each other, so one can obtain one from another by simply inverting one to get the other as described in the visual WOD. There is also one additional equation which fits quite well into the visual WOD, the Slutsky equation (called Slutsky in short in the visual WOD). Exhibiting the placement of the Slutsky equation in the WOD can contribute to deepening students’ understanding of its role in consumer theory. As depicted in the visual WOD, the Slutsky equation is an equation that is located between MDFs and HDFs, and relates the slopes of a MDF to its corresponding HDF. The Slutsky equation indeed attempts to explain changes in Marshallian demand due to changes in prices in terms of changes in Hicksian demand due to changes in prices and changes in Marshallian demand due to changes in income. That is, it is aimed at decomposing a price change into a substitution effect and an income effect. In essence, Slutsky equation attempts to give us some information on the duality gap between the solutions of the primal and dual problems; however, this equation shows this relationship in terms of partial derivatives, not absolute numerical values.

It is important to reiterate that the aim of the present paper is NOT to explain all the technical aspects of the components of the WOD, but rather to visually show how those numerous components are linked to each other. Therefore, the above explanation on the technical aspects of the WOD will suffice for the present purposes. Now, instead, turn your attention to the visual WOD itself.

In the following, six noteworthy points regarding some of the subtleties of the visual WOD are raised:

[8] It is important to differentiate between the pair of inverse MDFs and inverse HDFs (in which prices are expressed as functions of quantities) on the one hand, and the pair of HIDFs and AIDFs (in which “normalized” prices are expressed as functions of quantities) on the other hand. Although both of these pairs are inverse demand functions, the latter pair not only has inverse form but also has normalized prices.

- Alternatively, it is possible to set up the dual problem using the constraint $U(q) \geq \bar{u}$ instead of the constraint $D(q,u) = 1$, in which case the problem becomes as the following (figure 10):

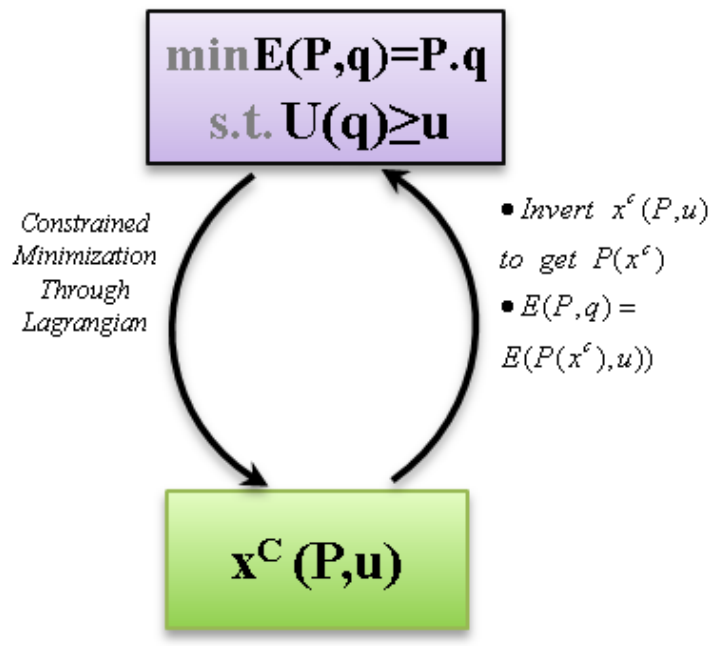

Figure 10. Alternative Setup for the Dual Problem in the WOD

where $E(P,q) = P \cdot q$ or in fact EAF is the objective function and $U(q) \geq \bar{u}$ is the constraint of a minimization problem. This way, however, it would not be possible to locate DF as a member of the quartet of functions representing preferences. Another consequence of using this version of the dual problem to get HDFs is that it will be inconvenient, though not impossible, to derive the system of AIDFs from it if we use this version of the dual problem in the WOD, since Antonelli equation has essentially been designed to turn DF into a system of AIDFs. Despite this, it pays to have this alternative way of deriving HDFs in mind, as it is in some cases a far more handy method to derive HDFs, especially when we do not need to know or derive DF nor AIDFs. In this case, it makes more sense to use the aforementioned dual problem instead of the one introduced in the visual WOD.

- A more detailed version of the lower-left portion of the WOD can be set up as the following (figure 11):

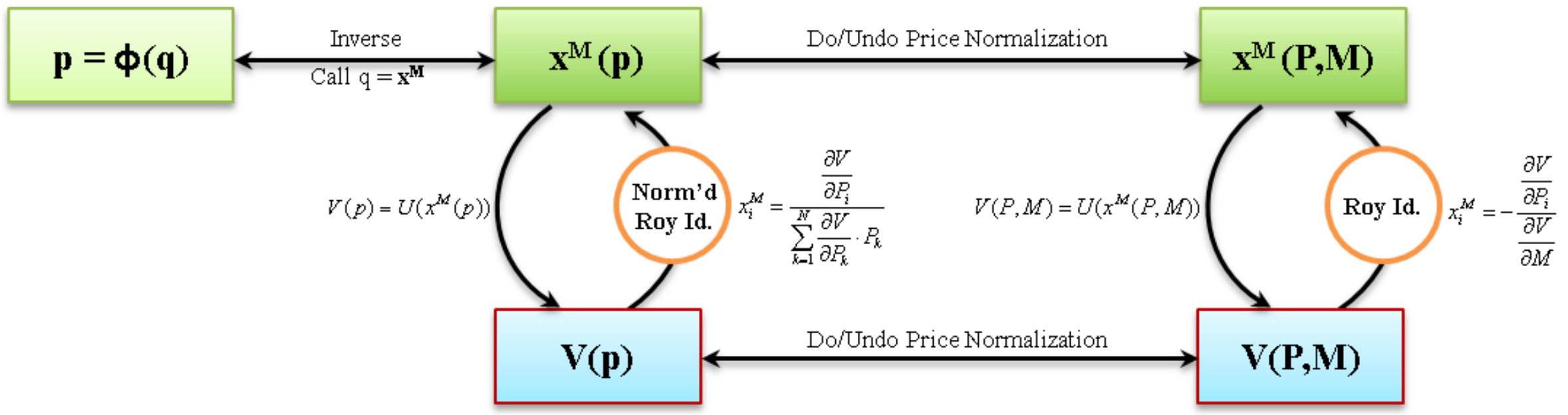

Figure 11. A More Detailed Version of the Lower-Left Portion of the WOD

where $x^M(p)$ is in fact $x^M(P/M, M/M) = x^M(p,1) = x^M(p)$ which is MDF with normalized prices, and $V(p)$ is indeed $V(P/M, M/M) = V(p,1) = V(p)$ which is IUF with normalized prices. As shown above, the transition from $V(p)$ to $x^M(p)$ is made through the normalized version of the Roy's identity (called Norm'd Roy Id. in short in the visual WOD). In this case, we can clearly see the placement and application of the Roy identity with normalized prices.

- A more detailed version of the lower-right portion of the WOD can be set up as the following (figure 12):

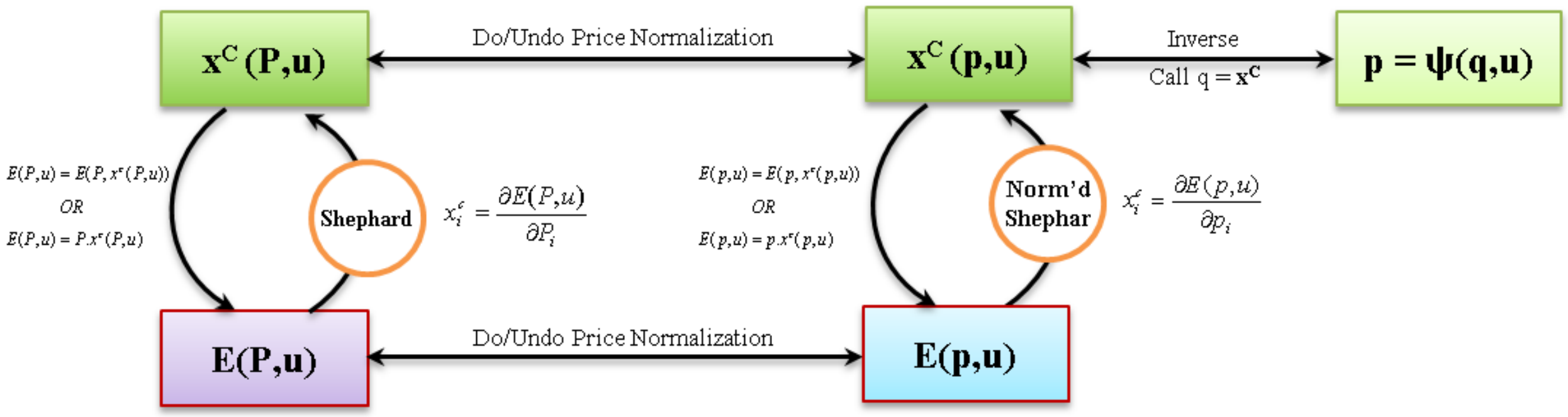

Figure 12. A More Detailed Version of the Lower-Right Portion of the WOD

where $x^C(p,u)$ is in fact $x^C(P/M, u) = x^C(p,u)$ which is HDF with normalized prices, and $E(p,u)$ is indeed $E(P/M, u) = E(p,u)$ which is EF with normalized prices. As shown above, the transition from $E(p,u)$ to $x^M(p,u)$ is made through the normalized version of the Shephard's Lemma (called Norm'd Shephard in short in the visual WOD).

- In transitioning from UF to MDFs, some information on preferences may be lost under some circumstances. As an example, if there is non-convexity in preferences, then there will be lost the non-convexity existing in the preferences when we make transition from DUF to MDFs. This loss of information occurs due to the maximization operation we use to derive a system of MDFs from DUF. Figure 13 illustrates this sort of information loss through a visual example.

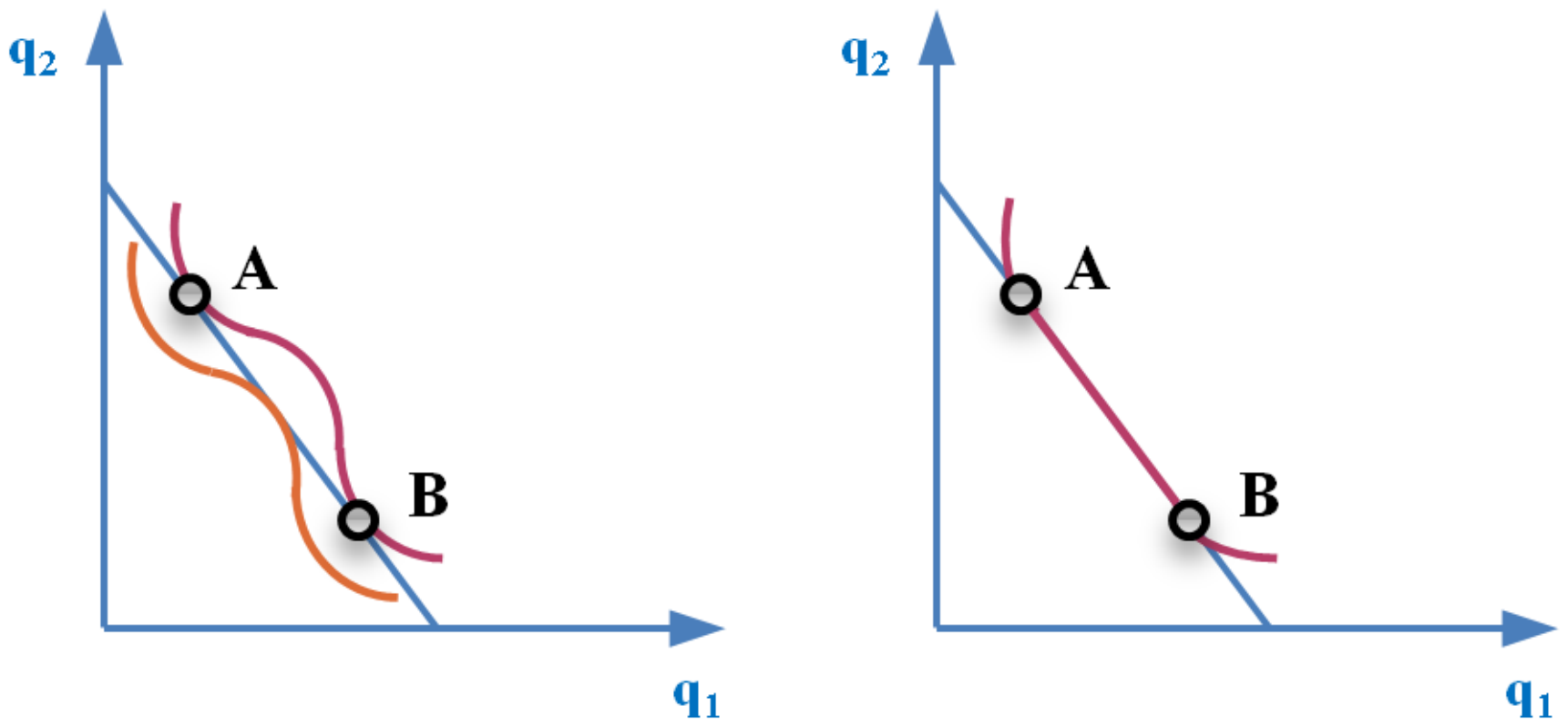

Figure 13. An Illustration Showing the Loss of Information on Preferences When Transitioning from DUF to MDFs

As is known in economics, agents in economic models are rational, meaning that they are goal-serving. In the present case, it is assumed that the consumer maximizes his or her utility function (as the objective function of the problem), during which procedure he or she chooses the combinations of the goods associated with the straight AB line on the right-hand side diagram, not the combinations associated with the curved AB line on the left-hand side diagram, which are all non-optimal.

Therefore, if there are some non-convex preferences similar to that of figure 13 on the left-hand side, and then if a system of MDFs is derived by solving a maximization problem, then consumers will automatically choose the combinations shown in figure 13 on the right-hand side, which does not involve any non-convexity anymore. Now, if we use the substitution technique introduced for transitioning from the system of MDFs to the DUF in order to make a complete loop, we will obtain the preferences demonstrated in figure 13 on the right-hand side, which is not identical to the original preferences (on the left hand side). The same loss of information naturally occurs in the non-convex preferences in the dual problem due to the minimization problem which is done to derive the system of HDFs.

- As a result of the explanation provided above, it must be crystal clear by now that there is a subtle difference between DUF and IUF in the sense that the latter embodies an optimizing process that the former does not (Cornes, 2008, p. 38). This is essentially because IUF is the result of the substitution of optimal quantities (MDFs) into the DUF, and optimal quantities do not reflect non-convexity and non-optimality. The same argument applies to DF and EF, meaning that the latter embodies an optimizing process that the former does not.
- Another interesting point to mention about the WOD is the fact that when there is no income effect involved (e.g. under quasi-linear preferences), MDF and HDF will coincide, meaning that we will obtain the same optimal solutions for both the primal and dual problems. In such cases, both MDF and HDF for a commodity are equally steep, and demand does not depend on income at all. Thus, there is no income effect on the demanded quantity of the commodity when prices change, so there will only be a substitution effect, and both MDF and HDF only show the substitution effect.

Speaking of pedagogical aspects of the visual WOD, it must be clear at this point that this graphic provides graduate economics students with a comprehensive visual "big picture" of the relationships among theoretical concepts in consumer theory. Nilson (2010) points out that "the younger generation of students is not as facile with text as it is with visuals, so a wise idea is to illustrate courses' designs to students so they can 'see' where the course is going in terms of students' learning." Visual aids such as graphic representation of theories, conceptual interrelationships, and knowledge

schemata – e.g. concept maps, mind maps, diagrams, flowcharts, comparison-and-contrast matrices, and the like – are powerful learning aids because they provide a ready-made, easy-to-process structure for knowledge (Svinicki, 2004; Vekiri, 2002).

Nilson (2010) believes instructors should give students the big picture – the overall organization of the course content – very early, and the clearest way to do this is in a graphic syllabus, and instructors should refer back to the visual big picture to show students how and where specific topics fit into that big picture (Nilson, 2010, p.242).[9] As Zeytoon Nejad (2016a) puts it, "if we, as instructors, take a course as consisting of three time phases, a big picture can help a class in all the three phases. In the first phase, it can be regarded as a graphical outline to illustrate where we are planning to go. In the middle phase, a "big picture" can be treated as a road map or a broad overview of the materials being covered in order to demonstrate exactly what and where in the course we are talking about at the moment. Thirdly, in the final phase, the big picture can be applied as a means of putting things together." Figure 13 depicts different roles that a visual "big picture" can play in different phases of a course.

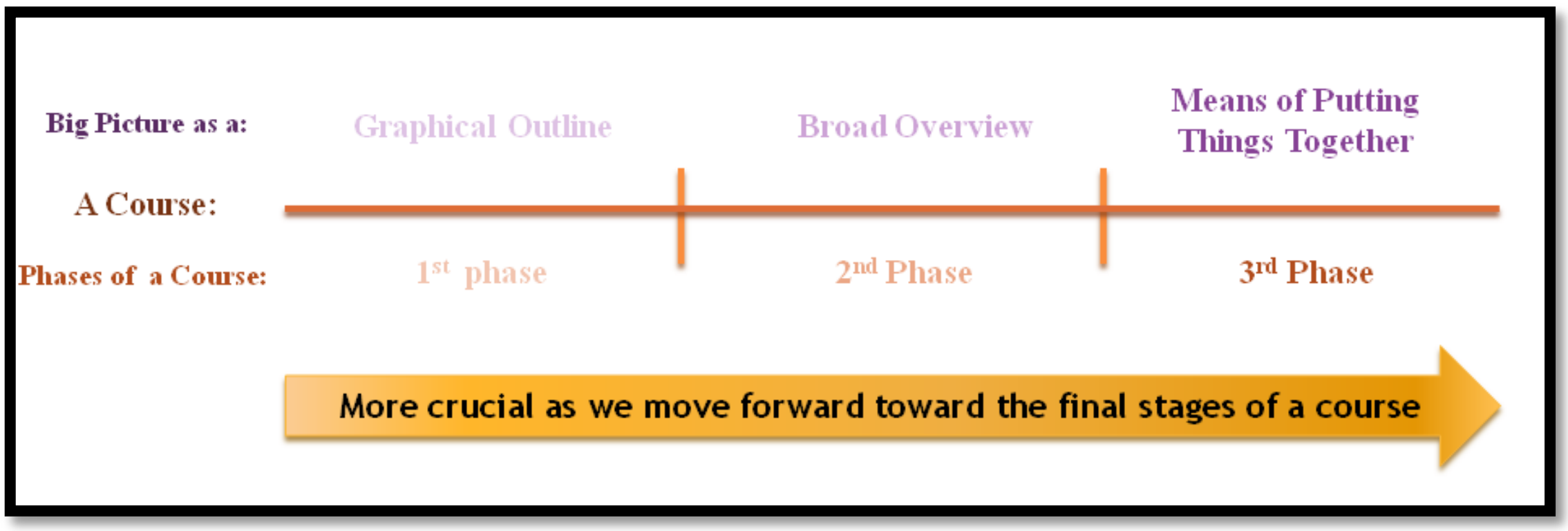

Figure 13. Different Uses of a Typical Big Picture in Various Stages of a Course

Having presented multiple papers with the same theme at different conferences, I have frequently received almost the same feedback from audiences stating that it would be wise to initially provide students with a crude version (a version excluding technical details and mathematical formulas) of a visual big picture in the first phase of a class. Instead, it would be nice to give students a complete version (a version including technical details and mathematical formulas) in the last phase of the class as a means of putting things together and wrap up the course in this way. I personally find this approach somewhat in line with the idea of "skeletal handout", which is usually discussed in the literature of education and teaching. This way, students will get more involved with the class. They will always seek to fill in the blanks by themselves. Additionally, by doing so, instructors give students a chance to fill in the blanks by their own words, signs, symbols, and according to their own learning styles and preferences. Finally, by giving them a complete, filled version of the visual big picture, you will let them correct their understanding of the subject matter at hand if they are mistaken.

As should be obvious, a typical visual "big picture" ignores a large amount of details; however, this is, in fact, its philosophy to do so. That is, the mission of a visual big picture, like the visual WOD introduced in this paper, is to retain the major ideas, and demonstrate the ways through which those major concepts are connected to each other. Therefore, a big picture serves as the framework of a course, and the lecture notes, lectures themselves, textbooks, and other sorts of the materials instructors typically take advantage of in classes will provide the needed details to deepen the students' understanding of the materials being covered (Zeytoon Nejad, 2016a).

## 4. Conclusion

Dual arguments, as standard tools of modern economic analyses, which are heavily involved with optimization, have been being used by many economists in recent decades. Despite this, a lack of simple, clear, and holistic explanations of the components of dual arguments is still disappointing. This paper made an effort to fill the said gap, with the hopes of helping economics students and economists deeply understand the duality theory applications in advanced microeconomics.[10] In other words, this paper graphically decoded the logical, complex relationships among a quartet of dual functions which represent preferences, namely DUF, IUF, DF, and EF as well as a quartet of demand functions, namely MDFs, HIDFs, HDFs, and AIDFs in a visual manner.

[9] Zeytoon Nejad (2016b) introduces a more innovative variation of syllabus called the Interactive Graphic Syllabus, and elaborates how one can design such an effective syllabus in the context of economics.

[10] The present paper introduced the visual wheel of duality for the case of consumer theory in modern microeconomics. Naumenko and Zeytoon Nejad Moosavian (2016) introduce the counterpart of this wheel of relationships for the case of producer theory. Also, Zeytoon Nejad Moosavian (2016d) introduces a somewhat similar idea for the case of intermediate macroeconomics.

In addition to the eight concepts mentioned above, the visual WOD presented in this paper introduced numerous other crucial microeconomic concepts, and explained in what ways these concepts are related to one another. Some of these concepts were Hotelling-Wold identity, Roy's identity, Shephard's lemma, Antonelli equation, Slutsky equation, budget constraint, expenditure amount function, among others.Afterwards, the paper brought up six noteworthy points about some of the subtleties of dual arguments in the context of consumer theory. In total, the comprehensive, visual WOD presented in this paper logically connected fourteen interrelatedly linked microeconomic concepts, and outlined how one can make sixteen microeconomically logical transitions among the aforementioned dual and demand functions.

The paper implicitly suggests that regardless of what courses economics instructors are teaching, they should not leave the structure they are building in their students' minds without a strong framework, which will be indeed their visual "big picture." Economics instructors can design their own visual "big pictures" according to their teaching experiences, personal preferences, ways of thinking, etc. They can also bring it up in different phases of their classes, whenever they prefer to do so. After all, what they should not do is leave their students without a "big picture" in their minds. Last but not least, the graphical demonstration of WOD could be a practical example of utilizing visualization to improve practices in the teaching of economics. It can also serve as an example supporting the notion that teaching and learning economics does not necessarily have to be difficult even at a graduate level.

**Acknowledgments**

I am very grateful to Dr. Walter Thurman for his valuable comments, suggestions, encouragement, and help in finding some of the missing parts of the WOD to me. Additionally, I would like to thank participants at the 80th International Atlantic Economic Conference held in Boston for their helpful discussions. Of course, all possible errors remain mine.

**References:**

Ahlheim, M. (1988). On the economics of the Antonelli equation. *European Journal of Political Economy, 4*(4), 539-552. http:/dx.doi.org/10.1016/0176-2680(88)90016-X

Anderson, J. R. (1993). Problem-solving and learning. *American Psychologist, 48*(1), 35-57. http:/dx.doi.org/ 10.1037/0003-066X.48.1.35

Arcavi, A. (2003). The role of visual representations in the learning of mathematics. *Educational Studies in Mathematics, 52***,** 215–241. http:/dx.doi.org/ 10.1023/A:1024312321077

Arocha, J. F., & Patel, V. L. (1995). Novice diagnostic reasoning in medicine: Accounting for clinical evidence. *Journal of the Learning Sciences, 4,* 355–384. Retrieved from http://www.jstor.org.prox.lib.ncsu.edu/stable/1466784

Boyd, S. P., & Vandenberghe, L. (2004). *Convex optimization* (pdf). Cambridge University Press. ISBN 978-0-521-83378-3.

Cornes, R. (2008). *Duality and modern economics,* Cambridge University Press.

Deaton, A. (1979). "The distance function in consumer behaviour with applications to index numbers and optimal taxation." *Review of Economic Studies, 46*(3), 391-405. Retrieved from http://www.jstor.org.prox.lib.ncsu.edu/stable/2297009

Gilbert, J. K. (2010). The role of visual representations in the learning and teaching of science: An introduction. *Asia-Pacific Forum on Science Learning and Teaching*, *11*(1), *Foreword,* 1. Retrieved from https://www.ied.edu.hk/apfslt/download/v11_issue1_files/foreword.pdf

Hotelling, H. (1932). "Edgeworth's taxation paradox and the nature of demand and supply," *Journal of Political Economy, 40,* 577–616. Retrieved from http://www.jstor.org/stable/1822600

Kozma, R. B., Russell, J., Jones, T., Marx, N., & Davis, J. (1996). The use of multiple linked representations to facilitate science understanding. In S. Vosniadou, E. DeCorte, R. Glaser, H. Mandl (Eds.), *International Perspectives on the Design of Technology – Supported Learning Environments* (41-60). Mahwah, NJ: Erlbaum.

Leemis, L., & McQueston, J. (2008). Univariate distribution relationships. *The American Statistician, 62*, 45-53. http:/dx.doi.org/ 10.1198/000313008X270448

McCormick, B. H., DeFantim, T. A., & Brown, M. D. (1987). Visualization in scientific computing: Definition, domain, and recommendations, *Computer Graphics*, *21*, 3–13.

Moosavian, S. A. Z. N. (2016). Teaching Economics and Providing Visual" Big Pictures". *arXiv preprint arXiv:1601.01771*. Retrieved from http://arxiv.org/abs/1601.01771

Moosavian, S. A. Z. N. (2016). Teaching Economics and Providing Visual "Big Pictures". *Journal of Economics and*

*Political Economy*, *3*(1), 119-133. http://dx.doi.org/10.1453/jepe.v3i1.631

Moosavian, S. A. Z. N. (2016c). A Comprehensive Visual "Wheel of Duality" in Consumer Theory. *International Advances in Economic Research*, 1-2. http:/dx.doi.org/ 10.1007/s11294-016-9586-8

Moosavian, S. A. Z. N. (2016d). The visual "big picture" of intermediate macroeconomics, Manuscript in preparation.

Naumenko, A., & Zeytoon, N. M. S. A. (2016). Clarifying theoretical intricacies through the use of conceptual visualization: Case of production theory in advanced microeconomics, Manuscript in preparation.

Naumenko, A., & Zeytoon, N. M. S. A. (2016, February). Clarifying theoretical intricacies through the use of conceptual visualization: Case of production theory in advanced microeconomics, Paper presented at the 27th Annual Teaching Economics Conference - Instruction and Classroom Based Research, held by Robert Morris University & McGraw Hill/Irwin Publishing Company, Pittsburgh, PA, USA.

Nilson, L. B. (2010). *Teaching at its best: A research-based resource for college instructors (3rd ed).* San Francisco: Jossey-Bass.

Royer, J., Cisero, C. A., & Carlo, M. S. (1993). Techniques and procedures for assessing cognitive skills. *Review of Educational Research, 63*, 210-224.

Shephard, R. W. (1953): *Cost and production functions*. Princeton University Press, Princeton NJ.

Snyder, C., & Nicholson, W. (2012). *Microeconomic theory: Basic principles and extensions*, *11th edition*, South-Western CENGAGE Learning.

Winston, W. (1994). *Operations research: Applications and algorithms* (3rd ed.). Belmont, Calif.: Duxbury Press.

Zeytoon, N. M. S. A. (2016, June). Employing Technology in Providing an Interactive, Visual "Big Picture" for Macroeconomics: A Major Step Forward towards the Web-Based, Interactive, and Graphic Syllabus, Paper presented at the Sixth Annual American Economic Association (AEA) Conference on Teaching and Research in Economic Education (CTREE), Atlanta, GA, USA.

Zeytoon, N. M. S. A. (2016b). Using the Interactive, Graphic Syllabus in the teaching of economics, Manuscript in preparation.

Zeytoon, N. M. S. A. (2016d). The visual "big picture" of intermediate macroeconomics, Manuscript in preparation.

Zeytoon, N. M. S. A. (2016a). Teaching Economics and Providing Visual "Big Pictures". *Journal of Economics and Political Economy*, *3*(1), 119-133. http://dx.doi.org/10.1453/jepe.v3i1.631

Zeytoon, N. M. S. A. (2016c). A comprehensive visual 'wheel of duality' in consumer theory, Research Note, International Advances in Economic Research. http:/dx.doi.org/ 10.1007/s11294-016-9586-8

Appendix 1. Symbols and Notations

**max:** Maximize

**min:** Minimize

**s.t.:** Subject to

**q:** Vector of Quantities Consumed

**P:** Vector of Prices

**M:** Income

**p:** Vector of Normalized Prices, i.e. P/M

**U(q):** Direct Utility Function (aka Utility Function)

**M ≥ P.q:** Budget Constraint

**V(P,M):** Indirect Utility Function

**V(p):** Indirect Utility Function with Normalized Prices

**E(P,u): "**The" Expenditure Function

**E(P,q):** The Amount of Expenditures

**D(q,u):** The Distance Function

**$x^M$ (P,M):** Marshallian (aka Uncompensated or Walrasian or Ordinary) Demand Function

**$x^M$ (p):** Vector of Normalized Marshallian Demand Function

**p=ϕ(q):** Vector of Hotelling-style Inverse Demand Function

**$x^C$ (P,u):** Vector of Hicksian (aka Compensated) Demand Function

**p=ψ(q,u):** Vector of Antonelli-style Inverse Demand Function

**H-W Id.:** Hotelling-Wold Identity

**Antonelli:** Antonelli Equation

**Slutsky:** Slutsky Equation

**Roy Id.:** Roy's Identity

**Norm'd Roy Id.**: Normalized Version of Roy's Identity

**Shephard:** Shephard's Lemma

**Norm'd Shephard:** Shephard's Lemma with Normalized Prices

**DUF:** Direct Utility Function

**IUF:** Indirect Utility Function

**EF:** Expenditure Function

**DF:** Distance Function

**HIDF:** Hotelling-style Inverse Demand Function

**MDF:** Marshallian Demand Function

**HDF:** Hicksian Demand Function

**AIDF:** Antonelli-style Inverse Demand Function

**EAF:** Expenditure Amount Function

**BC:** Budget Constraint

Appendix 2. Mathematical Formulas

**Substitution of MDF into DUF to Obtain the IUF:**

$$V(P,M) = U(x^M(P,M))$$

**IUF vs. EF:**

$$V(P,E(P,u)) = u \qquad AND \qquad E(P,V(P,M)) = M$$

**Shephard's Lemma:**

$$x_i^c = \frac{\partial E(P,u)}{\partial P_i}$$

**DUF vs. DF:**

$$U\left(\frac{q}{D(q,u)}\right) = u$$

**Hotelling-Wold Identity:**

$$\phi_i(q) = p_i = \frac{P_i}{M} = \frac{\frac{\partial U}{\partial q_i}}{\sum_{j=1}^{n}\frac{\partial U}{\partial q_j}.q_j}$$

**Antonelly Equation:**

$$\psi_i(q,u) = p_i = \frac{P_i}{M} = \frac{\partial D(q,u)}{\partial q_i}$$

**Slutsky Equation:**

$$\frac{\partial x_i^M(P,M)}{\partial P_j} = \frac{\partial x_i^c(P,u)}{\partial P_j} - \frac{\partial x_i^M(P,M)}{\partial M}.x_j(P,M)$$

**Substitution of HDF into EAF to Obtain the EF:**

$$E(P,u) = E(P,x^c(P,u))$$

$$OR$$

$$E(P,u) = P.x^c(P,u)$$

**Substitution of the Inverse of MDF into IUF to Obtain the DUF:**

$\bullet\ Invert\ \ x^M(P,M)$

$to\ \ get\ \ P(x^M)$

$\bullet\ U(q) =$

$V(P(x^M),M)$

**Substitution of the Inverse of HDF into EF to Obtain the EAF:**

$\bullet\ Invert\ \ x^c(P,u)$

$to\ \ get\ \ P(x^c)$

$\bullet\ E(P,q) =$

$E(P(x^c),u))$

**Roy's Identity:**

$$x_i^M = -\frac{\frac{\partial V}{\partial P_i}}{\frac{\partial V}{\partial M}}$$

**MDF vs. HDF:**

$$x^c = x^M(P,E(P,u))$$

$$x^M = x^c(P,V(P,M))$$

Appendix 3. A Larger Version of the "Wheel of Duality"

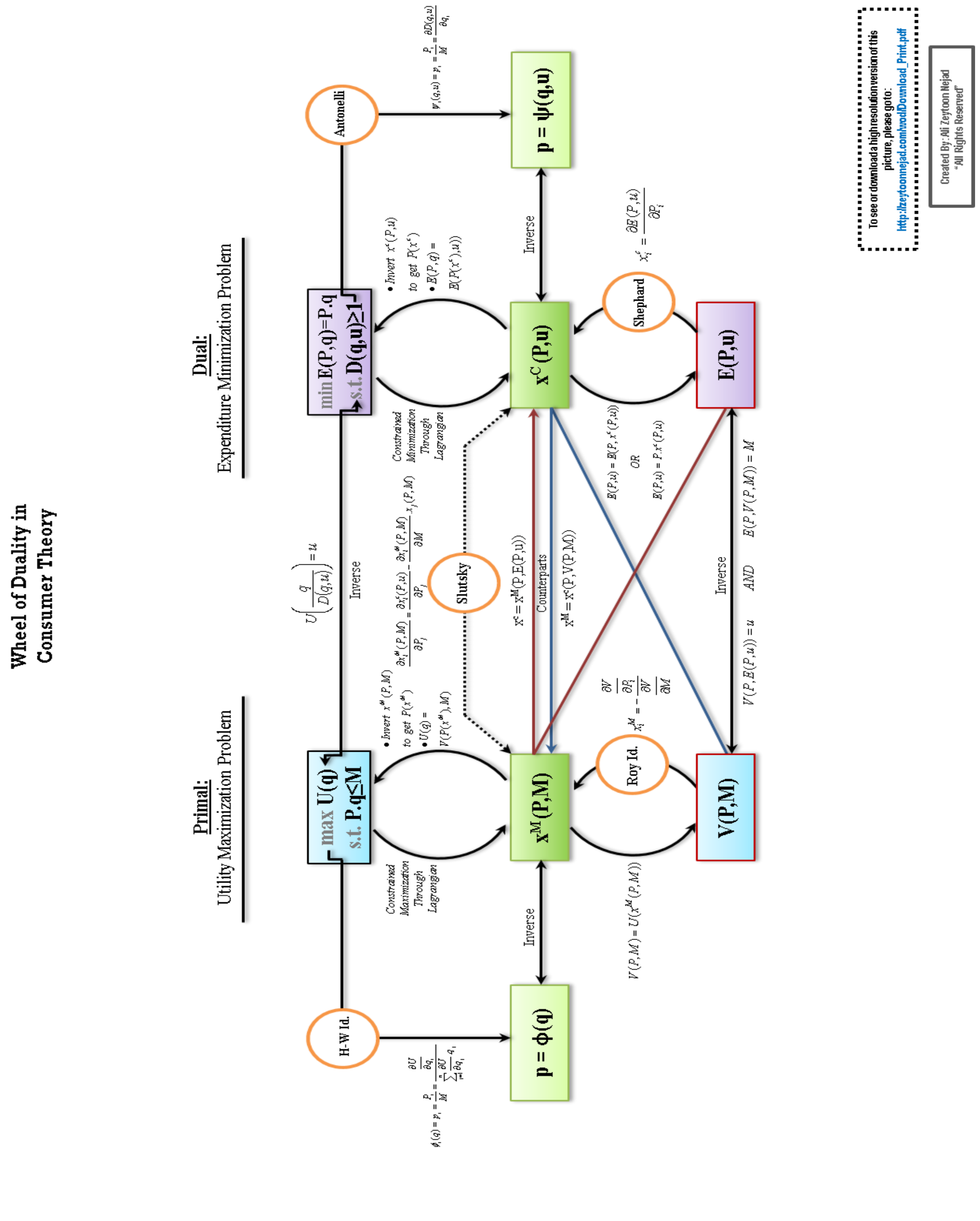